\documentclass[
    ,final            
  ]
  {aipproc}

\layoutstyle{6x9}

\begin{document}

\title{
Theoretical Description of K-Isomers }



\author{Yang Sun}{
address={Department of Physics and Joint Institute for Nuclear
Astrophysics, University of Notre Dame, Notre Dame, Indiana 46545,
USA} }



\begin{abstract}
A proper treatment of $K$-mixing is the key to understanding
$K$-isomers. Here, we present a method based on the projected
shell model. This method differs from the usual description of
multi-quasiparticle states by introducing a transformation to the
laboratory frame and a subsequent configuration mixing in that
frame. It allows a quantitative study on the degree of
$K$-violation through direct calculations of electromagnetic
transitions.
\end{abstract}

\maketitle


\section{Introduction}

Many long-lived, highly-excited isomers in deformed nuclei owe
their existence to the approximate conservation of the $K$ quantum
number \cite{BM,Walker1}.  The selection rule for an
electromagnetic transition would require that the multipolarity of
the decay radiation, $\lambda$, be at least as large as the change
in the $K$-value ($\lambda \ge \Delta K$).  However,
symmetry-breaking processes make possible transitions that violate
the $K$-selection rule; such `$K$-forbidden' transitions are
hindered, rather than strictly forbidden.  Much depending on the
degree of the $K$-violation, decays from a high $K$ state to
low-lying low $K$ states can have half-lives that range from
nanoseconds to years.

It is thus clear that a proper description of $K$-violation in
terms of $K$-mixing is at the heart of understanding $K$-isomers.
A theoretical model that can treat $K$-mixing has preferably the
basis states that are eigenstates of angular momentum $I$.
Diagonalization of two-body interactions mixes these states and
the resulting wavefunctions contain the information on the degree
of $K$-mixing.  In this kind of approach, the mixing and its
consequences are discussed in the laboratory frame rather than in
a body-fixed frame in which $K$ is originally defined.

\section{The projected shell model}

The projected shell model (PSM) \cite{PSM,PSMCode} is a shell
model that starts from a deformed basis.  In the standard version
of the PSM, the shell-model basis is constructed by considering a
few quasiparticle (qp) orbitals near the Fermi surfaces and
performing angular momentum projection (if necessary, also
particle-number projection) on the chosen configurations.  With
projected multi-qp states as the basis states of the model, the
PSM is designed to describe the rotational bands built upon qp
excitations \cite{PSM,PSM2}. The PSM has been rather successful in
calculating the high-spin states of normally-deformed and
superdeformed nuclei (see, for example, Refs.
\cite{PSM_SD190,Titled}).

In many of the existing calculations, the PSM begins with deformed
Nilsson single-particle states, with pairing correlations
incorporated into these states by a BCS calculation.  This defines
a set of deformed qp states (with $a^\dagger_\nu$ and
$a^\dagger_\pi$ being the creation operator for neutrons and
protons, respectively) with respect to the qp vacuum $|0\rangle$.
The PSM basis construction is then implemented in the multi-qp
states with the following forms
\begin{itemize}
\item e-e nucleus: ~~$\{ |0\rangle, a^\dagger_\nu a^\dagger_\nu
|0\rangle, a^\dagger_\pi a^\dagger_\pi |0\rangle, a^\dagger_\nu
a^\dagger_\nu a^\dagger_\pi a^\dagger_\pi |0\rangle, a^\dagger_\nu
a^\dagger_\nu a^\dagger_\nu a^\dagger_\nu |0\rangle, a^\dagger_\pi
a^\dagger_\pi a^\dagger_\pi a^\dagger_\pi |0\rangle, \ldots \}$

\item o-o nucleus: ~~$\{ a^\dagger_\nu a^\dagger_\pi |0\rangle,
a^\dagger_\nu a^\dagger_\nu a^\dagger_\nu a^\dagger_\pi |0\rangle,
a^\dagger_\nu a^\dagger_\pi a^\dagger_\pi a^\dagger_\pi |0\rangle,
a^\dagger_\nu a^\dagger_\nu a^\dagger_\nu a^\dagger_\pi
a^\dagger_\pi a^\dagger_\pi |0\rangle, \ldots \}$

\item odd-$\nu$ nucleus: ~~$\{ a^\dagger_\nu |0\rangle,
a^\dagger_\nu a^\dagger_\nu a^\dagger_\nu |0\rangle, a^\dagger_\nu
a^\dagger_\pi a^\dagger_\pi |0\rangle, a^\dagger_\nu a^\dagger_\nu
a^\dagger_\nu a^\dagger_\pi a^\dagger_\pi |0\rangle, \ldots \}$

\item odd-$\pi$ nucleus: ~~$\{ a^\dagger_\pi |0\rangle,
a^\dagger_\nu a^\dagger_\nu a^\dagger_\pi |0\rangle, a^\dagger_\pi
a^\dagger_\pi a^\dagger_\pi |0\rangle, a^\dagger_\nu a^\dagger_\nu
a^\dagger_\pi a^\dagger_\pi a^\dagger_\pi |0\rangle, \ldots \}$
\end{itemize}
The omitted index for each creation operator contains information
on the Nilsson orbitals.  In fact, this is the standard way of
building multi-qp states \cite{Jain95,Soloviev,Zeng02,Xu04}.
However, the present model goes a step beyond by transforming
these states from the body-fixed frame to the laboratory frame and
mixing them in the laboratory frame through two-body residual
interactions.

The angular-momentum-projected multi-qp states are thus the
building blocks in the PSM wavefunction, which can be generally
written as
\begin{equation}
|\psi^{I,\sigma}_M\rangle=\sum_{\kappa,K\le I}
f^{I,\sigma}_{\kappa} \hat P^{\,I}_{MK}|\phi_\kappa\rangle =
\sum_{\kappa} f^{I,\sigma}_{\kappa} \hat
P^{\,I}_{MK_\kappa}|\phi_\kappa\rangle .
\label{wavef}
\end{equation}
The index $\sigma$ labels states with same angular momentum and
$\kappa$ the basis states.  $\hat P^{\,I}_{MK}$ is the
angular-momentum-projection operator \cite{PSM} and the
coefficients $f^{I,\sigma}_{\kappa}$ are weights of the basis
states.

The weights $f^{I,\sigma}_{\kappa}$ are determined by
diagonalization of the Hamiltonian in the spaces spanned for
various nuclear systems as listed above, which leads to the
eigenvalue equation (for a given $I$)
\begin{equation}
\sum_{\kappa^\prime}\left(H_{\kappa\kappa^\prime}-E_\sigma
N_{\kappa\kappa^\prime}\right) f^\sigma_{\kappa^\prime} = 0.
\label{eigen}
\end{equation}
The Hamiltonian and the norm matrix elements in Eq. (\ref{eigen})
are given as
\begin{equation}
H_{\kappa\kappa^\prime}=\langle\phi_\kappa | \hat H \hat
P^I_{K_\kappa K^\prime_{\kappa^\prime}} | \phi_{\kappa^\prime}
\rangle , ~~~~~~~~~~ N_{\kappa\kappa^\prime}=\langle\phi_\kappa |
\hat P^I_{K_\kappa K^\prime_{\kappa^\prime}} |
\phi_{\kappa^\prime} \rangle .
\label{z}
\end{equation}
Angular-momentum-projection on a multi-qp state
$|\phi_\kappa\rangle$ with a sequence of $I$ generates a band. One
may define the rotational energy of a band (band energy) using the
expectation values of the Hamiltonian with respect to the
projected $|\phi_\kappa\rangle$
\begin{equation}
E^I_\kappa={H_{\kappa\kappa}\over
N_{\kappa\kappa}}={{\langle\phi_\kappa | \hat H \hat P^I_{K_\kappa
K_\kappa} | \phi_\kappa \rangle}\over {\langle\phi_\kappa | \hat
P^I_{K_\kappa K_\kappa} | \phi_\kappa \rangle}} .
\label{bande}
\end{equation}

In a usual approximation with independent quasiparticle motion,
the energy for a multi-qp state is simply taken as the sum of
those of single quasiparticles.  This is the dominant term.  The
present theory modifies this quantity in the following two steps.
First, the band energy defined in Eq. ({\ref{bande}) introduces
the correction brought by angular momentum projection and the
two-body interactions, which accounts for the couplings between
the rotating body and the quasiparticles in a quantum-mechanical
way. Second, the corresponding rotational states are mixed in the
subsequent procedure of solving the eigenvalue equation
({\ref{eigen}). The energies are thus further modified by the
configuration mixing.

If the deformed states are axially symmetric, each of the basis
states in (\ref{wavef}), the projected $|\phi_\kappa\rangle$, is a
$K$-state.  For example, an $n$-qp configuration gives rise to a
multiplet of $2^{n-1}$ states, with the total $K$ expressed by $K
= |K_1 \pm K_2 \pm \cdots \pm K_n|$, where $K_i$ is for an
individual neutron or proton.  In this case, shell model
diagonalization, i.e. solving the eigenvalue equation
(\ref{eigen}), is completely equivalent to $K$-mixing. The amount
of the mixing can be obtained from the resulting wavefunctions.

The above discussion is independent of the choice of the two-body
interactions in the Hamiltonian.  In practical calculations, the
PSM uses the pairing plus quadrupole-quadrupole Hamiltonian (that
has been known to be essential in nuclear structure calculations
\cite{Zuker96}) with inclusion of the quadrupole-pairing term
\begin{equation}
\hat H = \hat H_0 - {1 \over 2} \chi \sum_\mu \hat Q^\dagger_\mu
\hat Q^{}_\mu - G_M \hat P^\dagger \hat P - G_Q \sum_\mu \hat
P^\dagger_\mu\hat P^{}_\mu .
\label{hamham}
\end{equation}
The strength of the quadrupole-quadrupole force $\chi$ is
determined in such a way that it has a self-consistent relation
with the quadrupole deformation $\varepsilon_2$.  The
monopole-pairing force constants $G_M$ are
\begin{equation}
\begin{array}{c}
G_M = \left[ G_1 \mp G_2 \frac{N-Z}{A}\right] ~A^{-1} ,
\label{GMONO}
\end{array}
\end{equation}
with ``$-$" for neutrons and ``$+$" for protons, which reproduces
the observed odd--even mass differences in a given mass region if
$G_1$ and $G_2$ are properly chosen. Finally, the strength $G_Q$
for quadrupole pairing was simply assumed to be proportional to
$G_M$, with a proportionality constant 0.16, as commonly used in
the PSM calculations \cite{PSM}.

\section{Examples}

The nucleus $^{178}$Hf has become a center of recent discission on
$K$-isomers.  The possibility to trigger the 2.45MeV, 31-year
isomer decay by the application of external electromagnetic
radiation has attracted much interest and potentially could lead
to the controlled release of nuclear energy
\cite{Trigger,Walker2}.  Information on the detailed structure and
transition of the isomeric and the surrounding states thus becomes
a crucial issue.  In the PSM calculation for $^{178}$Hf
\cite{Hf178}, the model basis was built with the deformation
parameters $\varepsilon_2=0.251$ and $\varepsilon_4=0.056$ (values
taken from Ref. \cite{BFM86}).  Fig. 1 shows the calculated energy
levels in $^{178}$Hf, compared with the known data \cite{Mu97}.
Satisfactory agreement is achieved for most of the states, except
that for the bandhead of the first $8^-$ band and the $14^-$ band,
the theoretical values are too low.

It was found that the obtained states are generally $K$-mixed. One
may still talk about the dominant structure of each band by
studying the wavefunctions.  We found that the $6^+$ band has
mainly a 2-qp structure $\{ \nu [512]{5/2}^- \oplus \nu
[514]{7/2}^- \},$ the $16^+$ band has a 4-qp structure $\{ \nu
[514]{7/2}^- \oplus \nu [624]{9/2}^+ \oplus \pi [404]{7/2}^+
\oplus \pi [514]{9/2}^- \},$ the first (lower) $8^-$ band has a
2-qp structure $\{ \nu [514]{7/2}^- \oplus \nu [624]{9/2}^+ \},$
the second (higher) $8^-$ band has a 2-qp structure $\{ \pi
[404]{7/2}^+ \oplus \pi [514]{9/2}^- \},$ and the $14^-$ band has
a 4-qp structure $\{ \nu [512]{5/2}^- \oplus \nu [514]{7/2}^-
\oplus \pi [404]{7/2}^+ \oplus \pi [514]{9/2}^- \}.$   These
states, together with many other states (not shown in Fig. 1)
obtained from the same diagonalisation process, form a complete
spectrum including the high-$K$ isomeric states.  Electromagnetic
transitions between any two of these states can be directly
calculated by using the wavefunctions, as done in the previous
publications (see, for example, Refs. \cite{PSM3,PSM4}). Results
will be reported elsewhere.


\begin{figure}
\includegraphics[height=.4\textheight]{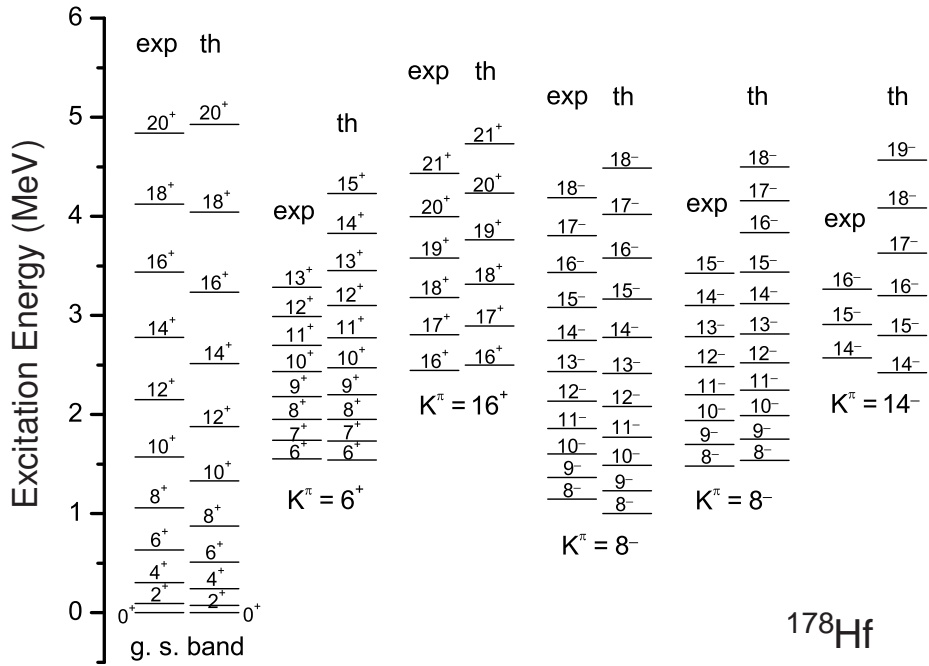}
\caption{Comparison of the PSM calculation for the rotational
bands in $^{178}$Hf with data.  This figure is adopted from Ref.
\cite{Hf178}.}
\end{figure}

\section{Further developments}

The preceding discussions suggest that the projected shell model
can be a useful tool for understanding $K$-isomers. So far, not
much has been done along these lines except for a few selected
applications \cite{Hf178,Nd134,Hf174}.  Therefore, there is a
large room for exploring the usefulness of the model. In addition,
three straightforward extensions may be necessary in order to
explain some of the current $K$-isomer data.
\begin{itemize}

\item \emph{Enlargement of the configuration space} -- There have
been impressive high-$K$ isomer data suggesting configurations
with up to 10-quasiparticles.  The current model space of the PSM
does not have these configurations and must be enlarged.

\item \emph{High-order multipole transitions} -- In some examples,
violation of the $K$-quantum number can only be understood through
electromagnetic transitions with high-order multipolarity.  The
current PSM code includes only the B(M1) and B(E2) parts.  The
high-order transition rates must be considered.

\item \emph{The effective forces} -- The pairing plus quadrupole
Hamiltonian has been proven successful for many of the structure
calculations.  However, other effects such as the octupole
correlation have also been considered important in some cases.
Additional terms, such as the monopole-monopole and
octupole-octupole interactions, can be included into the model if
these are important for certain discussions.

\end{itemize}

\section{Summary}

We have shown that the physics of $K$-mixing in multi-qp states is
well-incorporated in framework of the projected shell model. The
model follows the shell-model philosophy closely and may be viewed
as a shell model constructed in a projected multi-qp basis. More
precisely, the basis is first built in the qp basis with respect
to the deformed-BCS vacuum; then rotational symmetry, violated in
the deformed basis, is restored by angular-momentum-projection to
form a basis in the laboratory frame; finally a two-body
Hamiltonian is diagonalized in the projected multi-qp basis.  The
model can thus produce fully-correlated shell-model states and can
generate well-defined wavefunctions, allowing one to compute,
without any approximations, electromagnetic transition
probabilities.

\begin{theacknowledgments}
This work is partly supported by NSF under contract PHY-0140324.
\end{theacknowledgments}

\end{document}